\documentclass[aps,prl,twocolumn]{revtex4}
\usepackage{graphicx}
\usepackage{natbib}

\begin{document}

\title{On the origins of apparent fragile-to-strong transitions of protein hydration waters}

\author{M.\ Vogel}
\affiliation{Institut f\"ur Festk\"orperphysik, Technische Universit\"at Darmstadt,
Hochschulstr.\ 6, 64289 Darmstadt, Germany}

\begin{abstract}
$^2$H NMR is used to study the mechanisms for the reorientation of protein hydration
water. In the past, crossovers in temperature-dependent correlation times were reported at
$T_{x1}\!\approx\!225\mathrm{\,K}$ (X$_1$) and $T_{x2}\!\approx\!200\mathrm{\,K}$ (X$_2$). We show that neither X$_1$ nor X$_2$ are related to a fragile-to-strong transition. Our results rule out an existence of X$_1$. Also, they indicate that water performs thermally activated and distorted tetrahedral jumps at $T\!<\!T_{x2}$, implying that X$_2$ originates in an onset of this motion, which may be related to a universal defect diffusion in materials with defined hydrogen-bond networks.

\end{abstract}

\date{\today}

\maketitle

Studying the temperature dependence of the structural relaxation, one can distinguish
'strong' and 'fragile' glass-forming liquids, which do and do not show Arrhenius
behavior, respectively \cite{Angell_SCI_95}. For supercooled bulk water, it was proposed
that a fragile-to-strong transition (FST) exists at about $225\mathrm{\,K}$
\cite{Ito_NAT_99,Angell_SCI_08}. However, inevitable crystallization interferes with
direct observation \cite{Mishima_NAT_98}. In confinement, crystallization can be avoided
so that the dynamics of supercooled water are accessible down to the glass transition
temperature $T_g$. For confined water, crossovers in the $T$ dependence of a
correlation time $\tau$ were observed at $T_{x1}\!\approx225\mathrm{\,K}$ (X$_1$) and
$T_{x2}\!\approx200\mathrm{\,K}$ (X$_2$) using quasi-elastic neutron scattering (QENS)
\cite{Liu_PRL_05,Chen_PNAS_06} and dielectric spectroscopy (DS)
\cite{Bergman_NAT_00,Swenson_PRL_06}, respectively. Several workers took X$_1$ and X$_2$
as indications for a FST \cite{Liu_PRL_05,Chen_PNAS_06}. Also, X$_1$ was
related to a liquid-liquid phase transition
\cite{Liu_PRL_05,Chen_PNAS_06,Kumar_PRL_06,Zanotti_EPL_05}. Challenging these
conclusions, others argued that the Arrhenius processes P$_1$ and P$_2$ observed below X$_1$ and X$_2$, respectively, are secondary ($\beta$) relaxations
\cite{Cerveny_PRL_04,Sokolov_PRL_08}, while structural ($\alpha$) relaxation is difficult to observe \cite{Swenson_PRL_06}.

Confined waters are of enormous importance for biological, geological, and technological
processes, e.g., an interplay of water and protein dynamics enables biological functions
\cite{Fenimore_PNAS_04}. Here, we investigate the hydration waters of elastin (E) and collagen
(C), two proteins of the connective tissue. Using $^2$H NMR, we study time scale and mechanism of water reorientation during $T_{x1}$ and $T_{x2}$. The results show that a FST does not exist, while tetrahedral jumps become important upon cooling.

In $^2$H NMR, the quadrupolar frequency is probed \cite{Roessler}:
\begin{equation}\label{omega}
\omega_Q(\theta,\phi)=\pm\, \frac{\delta}{2}\,\left(3\cos^2\theta
-1-\eta\sin^2\theta\cos2\phi\right)
\end{equation}
Here, the angles ($\theta,\phi$) describe the orientation of the electric field gradient
(EFG) tensor at the nuclear site with respect to the external static magnetic field.
Since the molecular orientation determines the orientation of the EFG tensor, rotational
jumps render $\omega_Q$ time dependent. The anisotropy and asymmetry of the tensorial
interaction are characterized by $\delta$ and $\eta$, respectively. The $\pm$ signs
correspond to two allowed transitions between three Zeeman levels of the $^2$H nucleus
($I\!=\!1$).

C, E, and D$_2$O were purchased from Aldrich. Weighed amounts of protein and D$_2$O were
carefully mixed and sealed in the NMR tube to prepare samples with hydration levels
$h\!=\!0.25\!-\!0.96$ (g D$_2$O/$\mathrm{1\,g}$ protein). We refer to the samples as 'E'
or 'C' followed by the value of $h$ in percent. $^2$H NMR spin-lattice relaxation (SLR),
line-shape (LS), and stimulated-echo (STE) measurements are performed at a Larmor
frequency of $\omega_L\!=\!2\pi\!\times\!76.8\mathrm{\,MHz}$. The experimental setup is
described in Ref.\ \cite{Vogel_02}.

\begin{figure}
\begin{center}
\includegraphics[width=8.7cm]{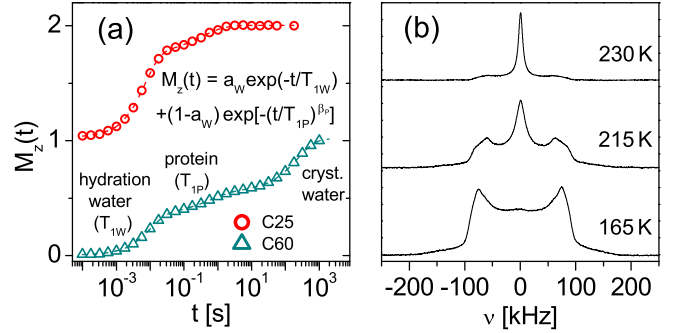}
\caption{(a)  Buildup of the magnetization $M_z(t)$ after saturation for C25 at $230\mathrm{\,K}$
(shifted) and C60 at $235\mathrm{\,K}$ together with fits to two- and three-step
relaxations, respectively. The two-step fitting function is indicated. (b) T-dependent
spectra of E43 resulting from the solid-echo sequence,
$90^\circ_x$--$\Delta$--$\,90^\circ_y$--$\Delta$--$\,t$. } \label{fig1}
\end{center}
\end{figure}

In the studied samples, supercoolable bound water and freezable free water can coexist.
Furthermore, deuterons replace exchangeable hydrogens of the protein \cite{NMR}. On the
basis of the amino acid compositions \cite{AS}, one expects 10--30\% of the deuterons to
be bound to proteins for the used hydration levels. Hence, deuterons of supercooled
water, crystalline water, and protein can contribute to $^2$H NMR signals. Accordingly,
three steps are observed in SLR measurements for C60, see Fig.\ \ref{fig1}(a). The third
step results from deuterons of crystalline water, as can be inferred from its absence for
low hydration levels, e.g., for C25, when free water does not exist
\cite{Nomura,Samou_02,Samou_04}. Since water deuterons outnumber protein deuterons, the
higher first and lower second steps can be assigned to supercooled water and protein,
respectively. The existence of distinct steps means that deuteron exchange between the
three species is slow on the time scale of SLR \cite{NMR}, enabling separate analysis of
the dynamical behaviors in $^2$H NMR.

\begin{figure}
\begin{center}
\includegraphics[width=8.7cm]{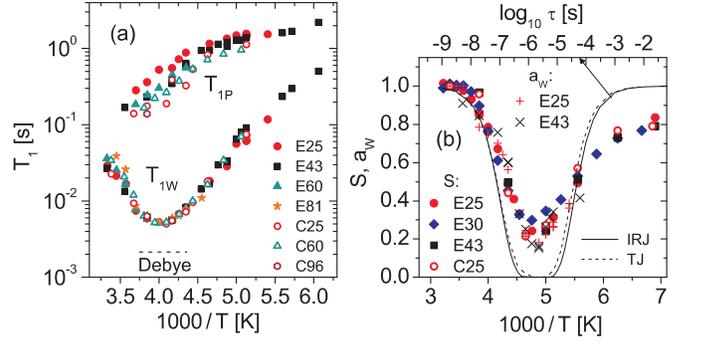}
\caption{(a) SLR times $T_{1w}$ and $T_{1p}$ for deuterons of supercooled waters and
proteins, respectively. The dashed horizontal line is the expectation for the $T_{1w}$
minimum in the case of a Debye process, as calculated from Eq.\ (\ref{T1}) for the
experimental values $\delta\!=\!2\pi\!\times\!168\mathrm{\,kHz}$ and
$\omega_L\!=\!2\pi\!\times\!76.8\mathrm{\,MHz}$. (b) T-dependent integrated intensity $S$
of solid-echo spectra and relative height $a_w$ of the first SLR step for hydrated E and
C samples. The lines show the dependence of $S$ on the correlation time $\tau$, as
obtained in simulations of isotropic random jumps (IRJ) and tetrahedral jumps (TJ), using
the experimental values $\delta\!=\!2\pi\!\times\!168\mathrm{\,kHz}$, $\eta\!=\!0.1$, and
$\Delta\!=\!20\mathrm{\,\mu s}$.} \label{fig2}
\end{center}
\end{figure}

Depending on the value of $h$, we use two- or three-step relaxations to fit the buildup
of magnetization, $M_z(t)$, see Fig.\ \ref{fig1}(a). While we will not discuss results
for crystalline water, Figure \ref{fig2}(a) compares the $T$-dependent SLR times $T_{1w}$
and $T_{1p}$ of deuterons in supercooled waters and proteins, respectively. $T_{1w}$
shows a very similar minimum for all samples, indicating that water dynamics in the
hydration shells of C and E are highly comparable. Unlike the first step, the second step
is nonexponential ($\beta_p\!\approx\!0.6$), typical of amorphous solids \cite{Roessler}.
For C and E, $T_{1p}$ and, thus, the protein dynamics weakly depend on $T$.

\begin{figure}
\begin{center}
\includegraphics[width=8.7cm]{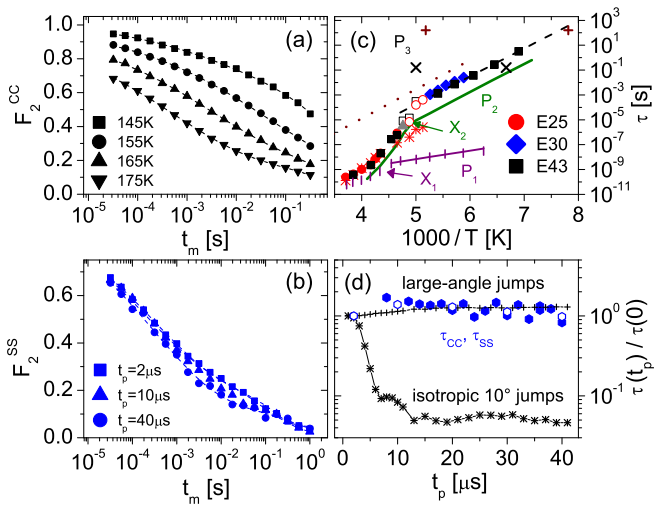}
\caption{(a) $F_2^{cc}(t_m;t_p\!=\!30\mathrm{\,\mu s})$ of E43 at various T and fits to
Eq.\ (\ref{f2fit}). (b) $F_2^{ss}(t_m;t_p\!=\!2,10,40\mathrm{\,\mu s})$ of E30 from PR experiments
at 185$\mathrm{\,K}$ and fits to Eq.\ (\ref{f2fit}). (c) Correlation times from LS (triangle)
and SLR analysis at $T\!\geq\!195\mathrm{\,K}$ ($J_{CD}$: circles and squares, $J_{CC}$: stars)  and from STE analysis below $\mathrm{195\,K}$ [E43:
$\langle \tau_{cc} \rangle$ from $F_2^{cc}(t_m;t_p\!=\!30\mathrm{\,\mu s})$, E30:
$\langle \tau_{ss} \rangle$ from PR $F_2^{ss}(t_m;t_p\!=\!2\mathrm{\mu
s})\!\approx\!F_2(t_m)$]. Open symbols mark differing SLR results from $J_{CD}$ and $J_{CC}$. Correlation times from measurements using DS for (solid line)
hydrated myoglobin \cite{Swenson_JPCM_07} and (dotted line) E ($h\!=\!0.1$) \cite{Samou_02}, ($+$) TSC
for E ($h\!=\!0.5$) \cite{Samou_04}, ($\times$) MR for C ($h\!\geq\!0.5$) \cite{Nomura},
and ($\mid$) QENS for hydrated lysozyme \cite{Chen_PNAS_06}. (d) Evolution-time dependence of $\langle \tau_{cc} \rangle$ (solid
circles) and $\langle \tau_{ss} \rangle$ (open circles) from PR STE experiments on E30 at
185$\mathrm{\,K}$ and simulation results for ($+$) distorted ($\pm 3^\circ$)
tetrahedral jumps and ($\ast$) isotropic $10^\circ$ jumps.} \label{fig3}
\end{center}
\end{figure}

Using that reorientation of the hydration waters is basically isotropic in the vicinity
of the minimum, see below, $T_{1w}$ depends on the spectral density according to
\cite{Roessler}
\begin{equation}\label{T1}
T_{1w}^{-1}=(2/15)\,\delta^2\left[J(\omega_L)+4J(2\omega_L)\right]
\end{equation}
Here, $J(\omega)\!=\!\int_0^\infty F_2(t)\cos(\omega t)dt$. The rotational correlation
function (RCF) $F_2(t)\!\propto\!\langle P_2[\cos\theta(0)] P_2[\cos\theta(t)]\rangle$
describes the time dependence of the Legendre polynomial $P_2$ of the angle $\theta$. For
a Debye process, $F_2(t)\!=\!\exp(-t/\tau)$ and
$J_{BPP}(\omega)\!=\!\tau/(1\!+\!\omega^2\tau^2)$. In this case, Eq.\ (\ref{T1}) takes
the form derived by Bloembergen, Purcell, and Pound (BPP) \cite{BPP}. However, this
approach predicts $T_{1w}\!=\!2.4\mathrm{\,ms}$ at the minimum, at variance with our
results in Fig.\ \ref{fig2}(a). Thus, a Debye process does not describe the water
dynamics, but a distribution of correlation times $G(\log \tau)$ exists, as typical of
supercooled liquids. To consider such distribution, Cole-Davidson (CD) and Cole-Cole (CC) spectral densities proved useful in SLR analyses
\cite{Roessler,Beckmann}
\begin{eqnarray}
J_{CD}(\omega)&=&\omega^{-1}\sin[\beta_{CD}\arctan(\omega\tau_{CD})]/(1\!+\!\omega^2\tau_{CD}^2)^{\frac{\beta_{CD}}{2}} \nonumber \\
J_{CC}(\omega)&=&\frac{\omega^{-1}\sin\left(\frac{\beta_{CC}\pi}{2}\right)(\omega\tau_{CC})^{\beta_{CC}}}{1+(\omega\tau_{CC})^{2\beta_{CC}}+2\cos\left(\frac{\beta_{CC}\pi}{2}\right)(\omega\tau_{CC})^{\beta_{CC}}} \nonumber
\end{eqnarray}
Employing $J_{CD}$ ($J_{CC}$), a width parameter $\beta_{CD}\!=\!0.22$ ($\beta_{CC}\!=\!0.50$) is obtained from $T_{1w}$ at the minimum. Assuming that $\beta_{CD}$ ($\beta_{CC}$) is independent of $T$ and inserting $J_{CD}$ ($J_{CC}$) into Eq.\ (\ref{T1}),
$\tau_{CD}$ ($\tau_{CC}$) is extracted from $T_{1w}$.  The mean correlation time
$\langle\tau_{CD}\rangle\!=\!\tau_{CD}\beta_{CD}$ \cite{Beckmann} of the asymmetric CD distribution and the peak position $\tau_{CC}$ of the symmetric CC distribution are shown for E25 and E43 in Fig.\ \ref{fig3}(c). At $220\!-\!260\mathrm{\,K}$, $J_{CD}$ and $J_{CC}$ yield comparable results, demonstrating insensitivity to the choice of the specific spectral density in this range. At lower T, non-Arrhenius and Arrhenius ($E_a\!=\!0.60\mathrm{\,eV}$) behaviors are obtained from use of $J_{CD}$ and $J_{CC}$,  respectively, so that ambiguity about $J(\omega)$ hampers analysis. Anyhow, neither $J_{CD}$ nor $J_{CC}$ yield evidence that X$_1$ exists.
 
In the following, we focus on $h\!\leq\!0.25\!-\!0.43$. Then, the hydration shells are
fully occupied, while freezable water hardly exists \cite{Nomura,Samou_04}. In Fig.\
\ref{fig1}(b), we see that the solid-echo spectra of E43 are comprised of two components.
Independent of T, the protein component is given by a broad spectrum, which is the
typical LS in the absence of motion \cite{Roessler}. For the water component, this LS is
observed at low T, but the broad spectrum collapses between 185 and $215\mathrm{\,K}$
resulting in a narrow Lorentzian at high T. The narrow spectrum reveals that, at
$T\!\geq\!215\mathrm{\,K}$, the water molecules show fast ($\tau\!\ll\!1/\delta$)
isotropic jumps that average out the anisotropy of the quadrupolar interaction, while
anisotropic motions, e.g., $\pi$ flips, do not result in a Lorentzian and can be
excluded. At
$T\!\leq\!185\mathrm{\,K}$, the broad spectrum indicates that significant water dynamics is absent on the time scale of $1/\delta\!\approx\!\mathrm{1\,\mu s}$. Thus, P$_1$, found in
QENS works below $T_{x1}$ \cite{Liu_PRL_05,Chen_PNAS_06}, is not the $\alpha$ process so that X$_1$ is not a FST. If P$_1$ were the $\alpha$ process, a narrow Lorentzian would be observed as $^2$H LS down to about $140\mathrm{\,K}$ due to $\tau_{P1}\!\ll\!1/\delta$, see Fig.\ \ref{fig3}(c), in clear
contrast to the findings in Fig.\ \ref{fig1}(b). The assignment and shape of the lines are confirmed when we single out the water contribution in partially relaxed (PR) experiments (not shown), in which we do not
wait for complete recovery of $M_z$ after saturation, but start acquisition at times between first and second SLR steps.

The total spectral intensity $S$, determined by integrating the solid-echo spectra after
correction for the Curie factor, exhibits a minimum $S\!\approx\!0.25$ at
$T\!\approx\!210\mathrm{\,K}$, see Fig.\ \ref{fig2}(b), indicating that water dynamics
during the dephasing and rephasing periods $\Delta$ of the solid-echo sequence interferes
with echo formation \cite{Schmidt}. Random-walk simulations \cite{Vogel_RW} for various
motional models show that $S$ is a minimum at $\tau\!=\!3\mathrm{\,\mu s}$. The signal is
not reduced for much faster (slower) dynamics, when $\omega_Q$ is
time-averaged (time-invariant). Deviations between measured and calculated data result because the simulations do not consider a distribution $G(\log \tau)$ and protein contributions. In Fig.\ \ref{fig3}(c), we see that LS and SLR
analyses yield consistent correlation times. In SLR
experiments, the height $a_w$ of the step due to supercooled water is also a minimum
at $210\mathrm{\,K}$ since a solid echo is used, see Fig.\ \ref{fig2}(b).

Finally, we
perform STE experiments \cite{Roessler} to study water dynamics at $T\!<\!T_{x2}$. In
STE experiments, we correlate the instantaneous quadrupolar frequencies $\omega_Q$ of a
deuteron during two short evolution times $t_p\!\ll\!\tau$ that are separated by a longer
mixing time $t_m\!\approx\tau$. Using appropriate pulse sequences, variation of $t_m$ for
constant $t_p$ enables measurement of RCF ($\xi\!=\!\mathrm{sin,cos}$; $x\!=\!ss,cc$)
\begin{equation}\label{SS}
F_2^{x}(t_m;t_p)\propto\langle\, \xi[\,\omega_Q(0)t_p]\,\xi[\,\omega_Q(t_m)t_p]\,\rangle
\end{equation}
The brackets $\langle\dots\rangle$ denote the ensemble average. $F_2^{ss}$ and $F_2^{cc}$
result for $\xi\!=\!\sin$ and $\xi\!=\!\cos$, respectively \cite{Roessler}. These RCF decay when slow
($t_p\!\approx\!10\mathrm{\,\mu s}\!\leq\!\tau\!\leq T_1\!\approx \! \mathrm{1\,s}$)
molecular reorientation alters the value of $\omega_Q(\theta,\phi)$ during $t_m$.

Figure \ref{fig3} presents $F_2^{x}(t_m;t_p)$ of E30 and E43. Water dynamics lead to
decays at short times ($\Phi_w$), while SLR results in additional damping of the water
($R_w$) and protein ($R_p$) contributions at longer times. Taking $R_w$ and $R_p$ from
the above SLR analysis, we fit $F_2^{x}(t_m)$ to
\begin{equation}\label{f2fit}
A_w[(1\!-\!B)\Phi_w(t_m)\!+B]R_w(t_m)\!+\!(1\!-\!A_w) R_p(t_m)
\end{equation}
Here, $B$ is introduced to consider that water dynamics does not destroy all
orientational correlation, see below. Using a stretched exponential
$\Phi_w(t_m)\!=\!\exp[-(t_m/\tau)^{\beta}]$, we obtain stretching parameters
$\beta\!=\!0.27\!-\!0.28$ \cite{BETA} for all studied values of $t_p$ and $T$. Thus,
hydration water exhibits strongly nonexponential RCF below $200\mathrm{\,K}$.  Use
of the $\Gamma$ function enables calculation of the mean correlation time, $\langle \tau
\rangle \!= \!(\tau/\beta) \Gamma(1/\beta)$. In Fig.\ \ref{fig3}(c), we see that $\langle
\tau_{ss} \rangle$ and $\langle \tau_{cc} \rangle$ from $F_2^{ss}$ of E30 and $F_2^{cc}$
of E43, respectively, follow the same Arrhenius law (dashed line) with activation energy
$E_a\!=\!0.45\mathrm{\,eV}$. We note that $F_2^{ss}$ of E30 was obtained in PR
experiments to minimize the protein contribution, leading to
$(1\!-\!A_w)\!=\!0.05\!-\!0.10$.

Comparison of our SLR, LS, and STE data implies a crossover in the $T$ dependence
of water dynamics in the vicinity of $T_{x2}$. Here, exact determination of a crossover temperature is hampered by a dependence on the choice of the spectral density used in SLR analysis. For an
understanding of the origin of X$_2$, knowledge about the mechanism for water dynamics
below $T_{x2}$ is of particular importance. We exploit that the dependence of STE
decays on the length of the evolution time has been shown \cite{Roessler,Fleischer} to
yield valuable insights into the geometry of rotational motion since sensitivity to small
changes of $\omega_Q$ and, thus, small angular displacements is higher for large $t_p$,
see Eq.\ (\ref{SS}). An analysis of $\langle \tau_{x} (t_p)\rangle$ enables a
determination of jump angles. While small-angle reorientation, results in a strong
decrease of $\langle \tau_{x} (t_p)\rangle$, $\langle \tau_{x} \rangle$ is independent of
$t_p$ for large-angle reorientation \cite{Roessler}, see Fig.\ \ref{fig3}(d). Extraction
of $\langle \tau_{x} (t_p)\rangle$ from $F_2^{x}(t_m;t_p)$ of E30 shows that a
substantial dependence on $t_p$ is absent, in harmony with results for C25 (not shown).
Hence, below $T_{x2}$, water exhibits large-angle rather than small-angle reorientation
typical of the $\alpha$ process \cite{Roessler}. In Fig.\ \ref{fig3}(b), we see that
$F_2^{ss}(t_m;t_p)$ decays to a small, but finite plateau $B(t_p)$, see Eq.\
(\ref{f2fit}), before the onset of SLR. The plateau height depends on the geometry of the
motion, e.g., $B(t_p\!\rightarrow\!\infty)\!=\!1/n$ for a $n$-site jump \cite{Fleischer}.
Here, $B$ decreases from 0.16 to 0.06 when $t_p$ is extended from 2 to $40\mathrm{\,\mu
s}$, indicating that, though water reorientation is not isotropic, angular restrictions
are not severe \cite{Fleischer}. While exact two- ($B\!=\!1/2$) or four-site
($B\!=\!1/4$) jumps can be excluded, distorted tetrahedral jumps, which may be expected
for a disordered hydrogen-bond network, are consistent with the data \cite{Fleischer}. In
$^2$H STE work on ice $I_h$ \cite{Fujara}, a similar dependence $B(t_p)$ was shown to
indicate that translational diffusion of water molecules via interstitial defects
involves distorted ($\pm 3^\circ$) large-angle reorientation between 7 O--D bond
orientations in the crystal.

In summary, we exploited the capabilities of $^2$H NMR to ascertain correlation times and mechanisms for the rotational motion of supercooled protein hydration waters. In the literature \cite{Bergman_NAT_00,Liu_PRL_05,Chen_PNAS_06,Kumar_PRL_06,Zanotti_EPL_05,Cerveny_PRL_04,Swenson_PRL_06,Sokolov_PRL_08}, it was a controversial issue to take crossovers X$_1$ and X$_2$ of temperature-dependent $\tau$ as evidence for a FST. Prerequisite for such interpretation of X$_1$ or X$_2$ is that the processes P$_1$ or P$_2$ below these crossovers are the $\alpha$ process. The present results, e.g., the LS at $145\!-\!185\mathrm{\,K}$, rule out that P$_1$ is the $\alpha$ process and, hence, that X$_1$ at $T_{x1}\!\approx\!\mathrm{225\,K}$ is a FST. While a narrow spectrum would result if P$_1$ were the $\alpha$ process, a broad spectrum is observed. Our findings conflict with QENS \cite{Liu_PRL_05,Chen_PNAS_06} and $^1$H NMR diffusion \cite{Mallamace} studies, attributing X$_1$ to a FST. Further evidence against X$_1$ being a FST comes from extrapolation of $\tau_{P1}(T)$, which yields too small values $T_g\!<\!\mathrm{100\,K}$. Also, a weak wave-vector dependence $\tau(Q)$ below X$_1$ \cite{Chen_PNAS_06} implies local rather than diffusive motion. Our results are consistent with the conjecture that P$_1$ is a $\beta$ process \cite{Cerveny_PRL_04,Swenson_PRL_06,Sokolov_PRL_08}, if the underlying motion is too restricted to be probed in $^2$H NMR.

Exploiting the sensitivity of $^2$H NMR to the mechanisms for water reorientation, we investigated the origin of X$_2$ at $T_{x2}\!\approx\!200\mathrm{\,K}$, reported in DS work \cite{Cerveny_PRL_04,Swenson_PRL_06}. Above $T_{x2}$, SLR and LS analyses showed that water exhibits isotropic reorientation described by a broad distribution $G(\log \tau)$. Below $T_{x2}$, STE experiments indicate that water performs large-angle jumps, most probably distorted tetrahedral jumps, which follow an Arrhenius law with $E_a\!=\!0.45\mathrm{\,eV}$. For P$_2$, DS work on water in various confinements \cite{Cerveny_PRL_04} reported comparable $\tau_{P2}(T)$, see Fig.\ \ref{fig3}(c). Also, mechanical-relaxation (MR) and thermally-stimulated currents (TSC) studies found a process with similar values of $\tau$ \cite{Nomura,Samou_04}. Therefore, all these methods probe P$_2$. However, P$_2$ neither destroys all orientational correlation, see Fig.\ \ref{fig3}, nor affects rigidity \cite{Nomura}. Thus, P$_2$ is not the $\alpha$ process and X$_2$ is not a FST. Moreover, the large-angle jump mechanism is strong evidence against P$_2$ being a Johari-Goldstein $\beta$ process \cite{Johari}, which results from small-amplitude reorientation \cite{Vogel_JG}. Comparison with previous $^2$H NMR work \cite{Fujara} showed that diffusion of water molecules via interstitial defects is consistent with the present results. Since $E_a\!\approx\!0.45\mathrm{\,eV}$ is not only found for water in various confinements \cite{Cerveny_PRL_04}, but also for crystalline and glassy bulk water \cite{Angell_SCI_08}, we suggest that water shows a characteristic tetrahedral jump motion, which is controlled by breaking of hydrogen bonds and, possibly, related to interstitial defect diffusion whenever a defined hydrogen-bond network is established, although we cannot exclude that the tetrahedral jump motion is governed by the protein surfaces in our case. The onset of this tetrahedral jump motion leads to more or less pronounced crossovers X$_2$. Extrapolating $\tau_{P2}(T)$, a value of $100\mathrm{\,s}$ is reached in the vicinity of $136\mathrm{\,K}$, the first widely accepted \cite{Mishima_NAT_98}, but later questioned \cite{Angell_SCI_08} value of $T_g$ for bulk water. Thus, one might speculate that the reported small calorimetric effects are not related to a glass transition, but to freezing of interstitial defects.

$^2$H NMR, which probes single-particle RCF, cannot resolve the issue whether the $\alpha$ process exists below X$_2$, because P$_2$ destroys basically all orientational correlation before an onset of structural relaxation. Detection of the $\alpha$ process rather requires techniques sensitive to the reorganisation of the whole hydrogen-bond network. In this respect, it is interesting that MR and TSC studies \cite{Nomura,Samou_04} found a slower process P$_3$,
see Fig.\ \ref{fig3}(c), which affects rigidity and, hence, may be the $\alpha$ process. Finally, it is an open question whether the onset of tetrahedral jump motion is related to a liquid-liquid phase transition, which was proposed to lead to a low-density liquid with a more ordered
tetrahedral network upon cooling \cite{Mishima_NAT_98,Angell_SCI_08}.

Funding of the DFG through grants VO 905/3-1 and VO 905/3-2 is gratefully acknowledged.


\begin{thebibliography}{99}

\bibitem{Angell_SCI_95} C.\ A.\ Angell, Science $\mathbf{267}$, 1924 (1995)
\bibitem{Ito_NAT_99} K.\ Ito, C.\ T.\ Moynihan, and C.\ A.\ Angell, Nature (London) $\mathbf{398}$, 492 (1999)
\bibitem{Angell_SCI_08} C.\ A.\ Angell, Science $\mathbf{319}$, 582 (2008)
\bibitem{Mishima_NAT_98} O.\ Mishima and H.\ E.\ Stanley, Nature (London) $\mathbf{396}$, 329 (1998)
\bibitem{Liu_PRL_05} L.\ Liu \emph{et al.}, Phys.\ Rev.\ Lett.\ $\mathbf{95}$, 117802 (2005)
\bibitem{Chen_PNAS_06} S.-H.\ Chen \emph{et al.}, Proc.\ Natl.\ Acad.\ Sci.\ $\mathbf{103}$, 9012 (2006)
\bibitem{Bergman_NAT_00} R.\ Bergman and J.\ Swenson, Nature (London) $\mathbf{403}$, 283 (2000)
\bibitem{Swenson_PRL_06} J.\ Swenson, H.\ Jansson, and R.\ Bergman, Phys.\ Rev.\ Lett.\ $\mathbf{96}$, 247802 (2006)
\bibitem{Kumar_PRL_06} P.\ Kumar \emph{et al.}, Phys.\ Rev.\ Lett.\ $\mathbf{97}$, 177802 (2006)
\bibitem{Zanotti_EPL_05} J.-M.\ Zanotti, M.-C.\ Bellissent-Funel, and C.-H.\ Chen, Europhys.\ Lett.\ $\mathbf{71}$, 91 (2005)
\bibitem{Cerveny_PRL_04} S.\ Cerveny \emph{et al.}, Phys.\ Rev.\ Lett.\ $\mathbf{93}$, 245702 (2004)
\bibitem{Sokolov_PRL_08} S.\ Pawlus, S.\ Khodadadi, and A.\ P.\ Sokolov, Phys.\ Rev.\ Lett.\ $\mathbf{100}$, 108103 (2008)
%\bibitem{Rupley_APC_91} J.\ A.\ Rupley and G.\ Careri, Adv.\ Protein Chem.\ $\mathbf{41}$, 37 (1991)
\bibitem{Fenimore_PNAS_04} P.\ W.\ Fenimore \emph{et al.}, Proc.\ Natl.\ Acad.\ Sci.\ $\mathbf{101}$, 14408 (2004)
%\bibitem{Doster_NAT_89} W.\ Doster, S.\ Cusack, and W.\ Petry, Nature (London), $\mathbf{337}$, 754 (1989)
%\bibitem{Rasmussen_NAT_92} B.\ F.\ Rasmussen \emph{et al.}, Nature (London), $\mathbf{357}$, 423 (1992)
%\bibitem{Spiess} K.\ Schmidt-Rohr and H.\ W.\ Spiess, \emph{Multidimensional Solid-State NMR and Polymers} (Academic Press, London, 1994)
\bibitem{Roessler} R.\ B\"ohmer \emph{et al.}, Prog.\ Nucl.\ Magn.\ Reson.\ Spectrosc.\ $\mathbf{39}$, 191 (2001)
\bibitem{Vogel_02} M.\ Vogel and T.\ Torbr\"ugge, J.\ Chem.\ Phys.\ $\mathbf{125}$, 164910 (2006)
\bibitem{NMR} G.\ E.\ Ellis and K.\ J.\ Packer, Biopolymers $\mathbf{15}$, 813 (1976)
\bibitem{AS} N.\ Vyavahare \emph{et al.}, Am.\ J.\ Pathology $\mathbf{155}$, 973 (1999)
\bibitem{Nomura} S.\ Nomura \emph{et al.}, Biopolymers $\mathbf{16}$, 231 (1977)
\bibitem{Samou_02} V.\ Samouillan \emph{et al.}, Biomacromolecules $\mathbf{3}$, 531 (2002)
\bibitem{Samou_04} V.\ Samouillan \emph{et al.}, Biomacromolecules $\mathbf{5}$, 958 (2004)

\bibitem{BPP} N.\ Bloembergen, E.\ M.\ Purcell and R.\ V.\ Pound, Phys.\ Rev.\ 73, 679 (1948)
\bibitem{Beckmann} C.\ P.\ Lindsey and G.\ D.\ Patterson, J.\ Chem.\ Phys.\ 73, 3348 (1980); P.\ A.\ Beckmann, Phys.\ Rep.\ $\mathbf{171}$, 85 (1988)
\bibitem{Schmidt} C.\ Schmidt, K.\ J.\ Kuhn, and H.\ W.\ Spiess, Colloid \& Polymer Sci.\ $\mathbf{71}$, 71 (1985)
\bibitem{Vogel_RW} M.\ Vogel and E.\ R\"ossler, J.\ Magn.\ Reson.\ $\mathbf{43}$, 147 (2000)
\bibitem{BETA} Using $\beta\!=\!0.97\beta_{CD}\!+\!0.14$, see Ref.\ \cite{Beckmann}, $\beta_{CD}\!=\!0.22$ from SLR
analysis translates into $\beta\!=\!0.35$.
\bibitem{Fleischer} G.\ Fleischer and F.\ Fujara, \emph{NMR - Basic Principles and Progress} (Springer, Berlin, 1994) Vol.\ 30, p. 159
\bibitem{Fujara} B. Geil, T.\ M.\ Kirschgen, and F.\ Fujara, Phys.\ Rev.\ B $\mathbf{72}$, 014304 (2005)
\bibitem{Mallamace} F.\ Mallamace \emph{et al.}, J.\ Chem.\ Phys.\ $\mathbf{127}$, 045104 (2007)
\bibitem{Johari} G.\ P.\ Johari and M.\ Goldstein, J.\ Chem.\ Phys. $\mathbf{53}$, 2372 (1970)
\bibitem{Vogel_JG} M.\ Vogel and E.\ R\"ossler, J.\ Chem.\ Phys.\ $\mathbf{114}$, 5802 (2001)
\bibitem{Swenson_JPCM_07} J.\ Swenson \emph{et al.}, J.\ Phys.: Condens.\ Matt.\ $\mathbf{19}$, 205109, (2007)
\end{thebibliography}
\end{document}